\documentclass[aps,prl,showpacs,floats,twocolumn,amsfonts,amssymb,unsortedaddress,floatfix]{revtex4}
\input epsf
\usepackage{bm}
\usepackage{amsmath}
\usepackage{graphicx}

\begin{document}

\addtolength{\topmargin}{10pt}

\def\Bbb{\mathbb}

\title{ Topological Solitons in Helical Strings}

\author{Cristiano Nisoli$^{1}$ and Alexander V. Balatsky$^{1,2}$}
\affiliation{\mbox{$^{1}$Theoretical Division and Institute for Material Sciences, Los Alamos National Laboratory, Los Alamos NM 87545 USA} \\
\mbox{$^{2}$Nordita, KTH Royal Institute of Technology and Stockholm University,  Stockholm, Sweden}}

%\date{\today}
\begin{abstract}
The low-energy  physics of  (quasi)degenerate one-dimensional systems is typically understood as the particle-like dynamics of kinks between stable, ordered structures. Such dynamics, we show, becomes highly non-trivial  when the ground states are   topologically constrained:  a dynamics {\it of the domains} rather than {\it on the domains} which the kinks separate.  Motivated by   recently reported observations of charged polymers physisorbed on  nanotubes,  we study  kinks between helical structures of a string wrapping  around a cylinder. While their motion cannot be disentangled from domain dynamics, and energy and momentum is not concentrated in the solitons, the dynamics of the domains can be folded back into a one-particle picture. 
\end{abstract}


\maketitle

The relationship between topological and physical properties~\cite{Shapere, Nelson, Lubensky} has received much recent attention. It is relevant  to elasticity~\cite{Landau, Lubensky}, non-linear physics~\cite{Dauxois,Ablowitz}, soft and hard  condensed matter~\cite{Nelson, Lubensky, Kamien}, and quantum computing~\cite{Quantum1, Quantum2}. %As topology is the study of invariance under  homeomorphism, it  shines a light on continuum field theories. 
 Topological invariants associated to physical objects often dictate  interaction: for instance  punctures in a plane (defects, dislocations, vortices) define a topological invariant (the winding angle)  and thus a logarithmic field which non surprisingly also to mediates their mutual interaction~\cite{Landau}. Similarly,  topologically distinct states  support infinitely continuum transitions~\cite{Kosterlitz, Nisoli14}. 
%Consider e.g. topological defects in 2D (punctures of the plane representing dislocations, vortices et cetera) whose interaction is mediated by logarithmic green functions, the same which defines the winding 
%The simplest topological objects of physical relevance is are punctures (defects) in a plane. They naturally defines topological invariants (the winding angle) from which  classical or quantum states of different helicity are topologically protected. From it comes a logarithmic field---which non surprisingly is   the (elastic) green function mediating  interactions between  such defects  on a plane []: This is only one example  of physical interactions dictated by  topological invariants []. Similarly, we learnt that systems of topologically distinct states can support infinitely continuum transitions []. 

We have previously investigated~\cite{Nisoli14} the statistical mechanics (and  connections with conformal invariance in quantum mechanics)  of  topological  transitions among  winding states representing   winding/unwinding polymers. Here we study the Newton dynamics of a self-interacting string (polymer) winding around a cylinder (nanotube). If strings are stable in  different, and non necessarily degenerate, helical structures,  they exhibit topological solitons  whose dynamics, however, is not ``contained'' in the kink but  involve the entire system. This is a feature of the topology of  helical solitons  found also in systems of essentially different physics: in   ``dynamical phyllotaxis''~\cite{NisoliP1, NisoliP2}   repulsive particles in cylindrical geometries mimic botanical patterns of leaves on stems, spines on cactuses, petals on a flower~\cite{Adler} by  self-organizing in helical lattices described by Fibonacci numbers~\cite{Levitov, NisoliP1}, also separated by kinks; or in colloidal crystals on cylinders and rod-shaped bacterial cell walls~\cite{Amir}.

While our analysis elucidates an interesting case of topology-dictated dynamics connected to the simplest topological invariant---the winding number---it is not without practical  implications. Polymer-nanotube hybrids, ssDNA-carbon nanotubes in particular~\cite{Williams, Couet, Zheng, Zheng2}, have been the subject of much recent experimental  and numerical research~\cite{Williams, Couet, Zheng, Zheng2, Johnson3, Johnson, Gigliotti, Johnson2, Manohar, Yarotski, Kilina} as promising candidates for  nanotechnological applications in bio-molecular and chemical sensing, drug delivery~\cite{Williams, Gannon} and dispersion/patterning of carbon nanotubes~\cite{Zheng, Zheng2, Gigliotti}. Indeed, ss-DNA forms tight helices on carbon nanotubes after sonication of the hybrids, although the role of base dependance and nanotube chirality is still debated~\cite{Zheng2, Gigliotti,Yarotski}, and raises  issues about how long-range order is reached, %One might speculate an analogy between such sonication and vibrofluidization in granular systems~\cite{Jaeger, Behringer, Abate, Sollich,Colizza, Song, Cugliandolo, Metha, Makse, Danna,Wang,NisoliD,NisoliT, Cugliandolo2}---or magneto agitation for magnetic materials~\cite{Wang,NisoliD,NisoliT}, which has been shown to be docile  to   descriptions in terms of an effective temperature~\cite{NisoliD,NisoliT, Cugliandolo2}.
not impossibly via an out-of-equilibrium phase transition~\cite{Evans} in a 1-D system with long range interactions~\cite{Fisher}. Order could then can come from interacting kinks  driven to coalesce and annihilate. 

 Theoretical research has so far concentrated on the chemical physics of the DNA-nanotube interaction~\cite{Johnson3, Johnson, Johnson2} and structure of the adsorbed polymer~\cite{Kilina} as well as on coarse grained modeling of the hybrid~\cite{Manohar}. However we know of no physics-based analysis rooted in the topology of the problem. 
We provide it here by describing the low-energy physics of these systems in terms of  the Newtonian dynamics of their kinks---implications for an over-damped,  driven regime will be reported elsewhere~\cite{NisoliFuture}. 
With a minimal, mesoscale, continuum model (M1),  we conceptualize the statics and low-energy nonlinear dynamics of a charged polymer physisorbed on a nanotube. Conclusions are corroborated by numerical analysis of a more faithful ball-and-spring model (M2).

We start with M1. Consider a 2D field $ {\boldsymbol \psi}(s,t)$ which describes a string (polymer) constrained to the surface of a cylinder or radius $r$: in cylindrical coordinates $z,r,\theta$ we have  $\psi_1=z$, $\psi_2=r\theta$ (see Fig. 1). $s$ is the intrinsic coordinate of the string, and as such, ${\boldsymbol T}={\boldsymbol \psi}'$ is its tangent vector  in the space $r \theta,~ z$. We write for $ {\boldsymbol \psi}(s,t)$ the following density of Lagrangian
\begin{equation}
{\cal L}= \frac{1}{2}\lambda \dot{{\boldsymbol \psi}}^2-\frac{1}{2}k {\left(\partial_s \boldsymbol T \right)}^2-V({\boldsymbol T})
\label{L}
\end{equation}
where $\lambda$ is the linear density of mass, $k$ of {\it effective} bending rigidity, and $V$ of an energy that depends only on the tangent vector (we denote  time derivatives with a dot): we thus assume that the long wavelength dynamics of polymers provides an effective smoothened potential, which affords  an analytical analysis. In a more realistic set-up, a site dependent potential will be used, a ball-and-spring model (M2) of a self interacting polymer of which M1 is the continuum generalization. %We will see then that the last two terms in Eq.~(\ref{L}) come from self-interaction screened by the tube.
Naturally, $V$ contains the possible symmetry breaking of the chiral structure and has the form of a double dip providing  two stable helices,  oppositely winding  and degenerate (Fig. 1, bottom left).  In general (e.g. because of a   corrugation potential of the tube) $V$ can have non degenerate {\it local} minima corresponding to different (meta)stable helices (Fig. 1, bottom right). 

\begin{figure}[t!]   
\includegraphics[width=3.2 in]{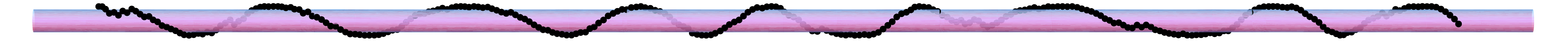}\vspace{2mm}
\includegraphics[width=3.1 in]{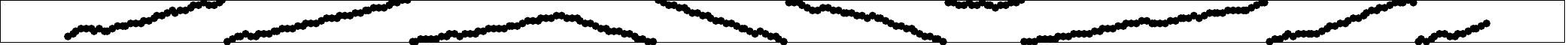}\vspace{2mm}
\includegraphics[width=3.25 in]{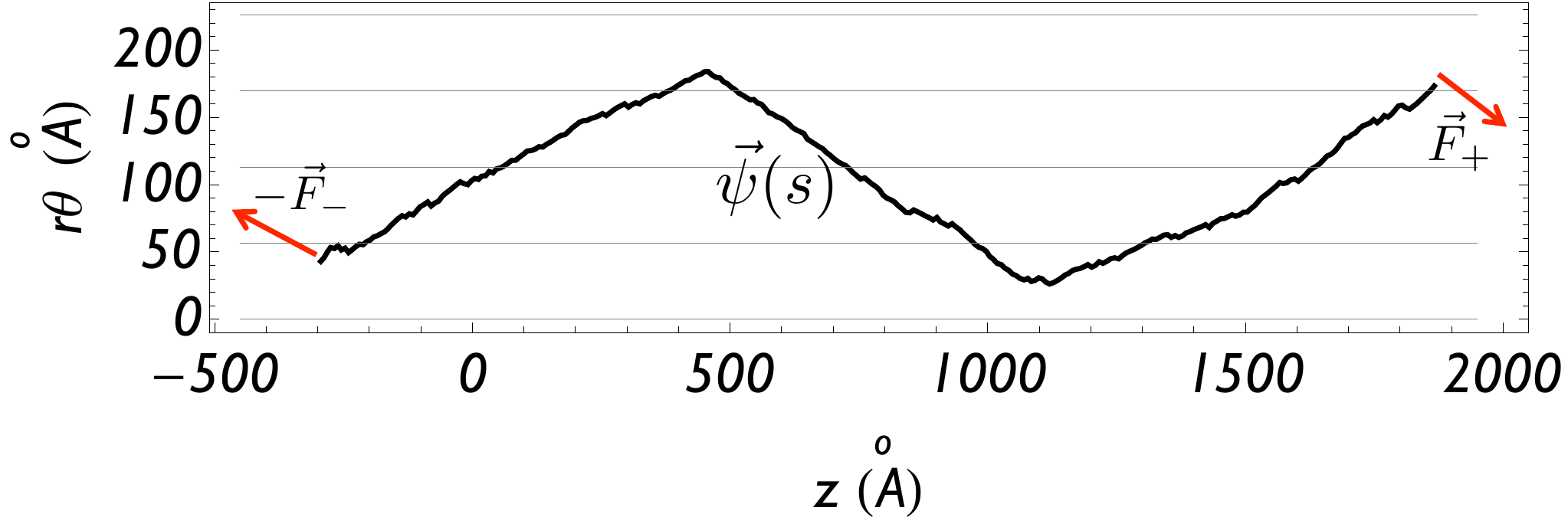}\vspace{4mm}
\includegraphics[width=1.5 in]{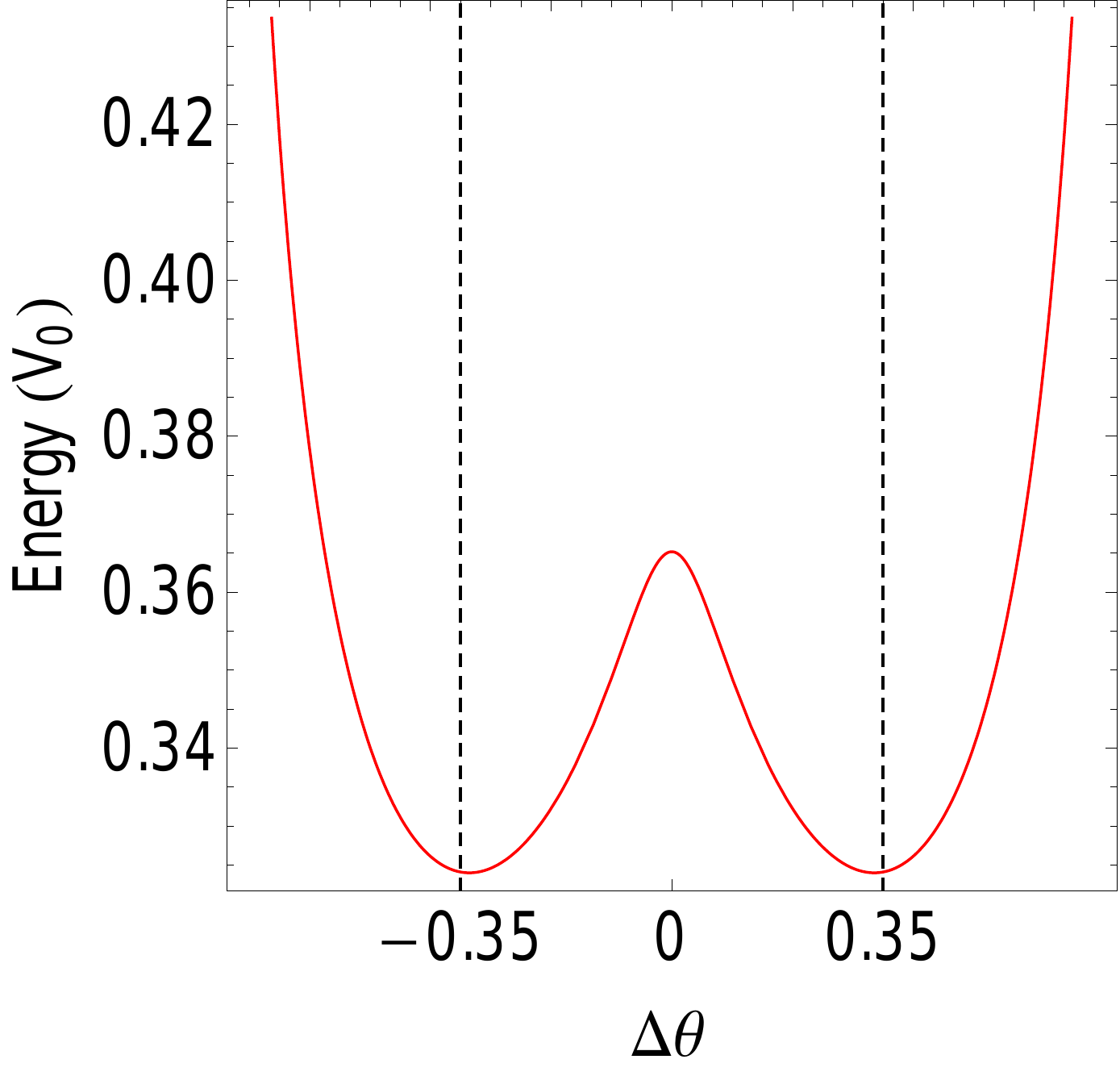}\hspace{5mm}\includegraphics[width=1.5 in]{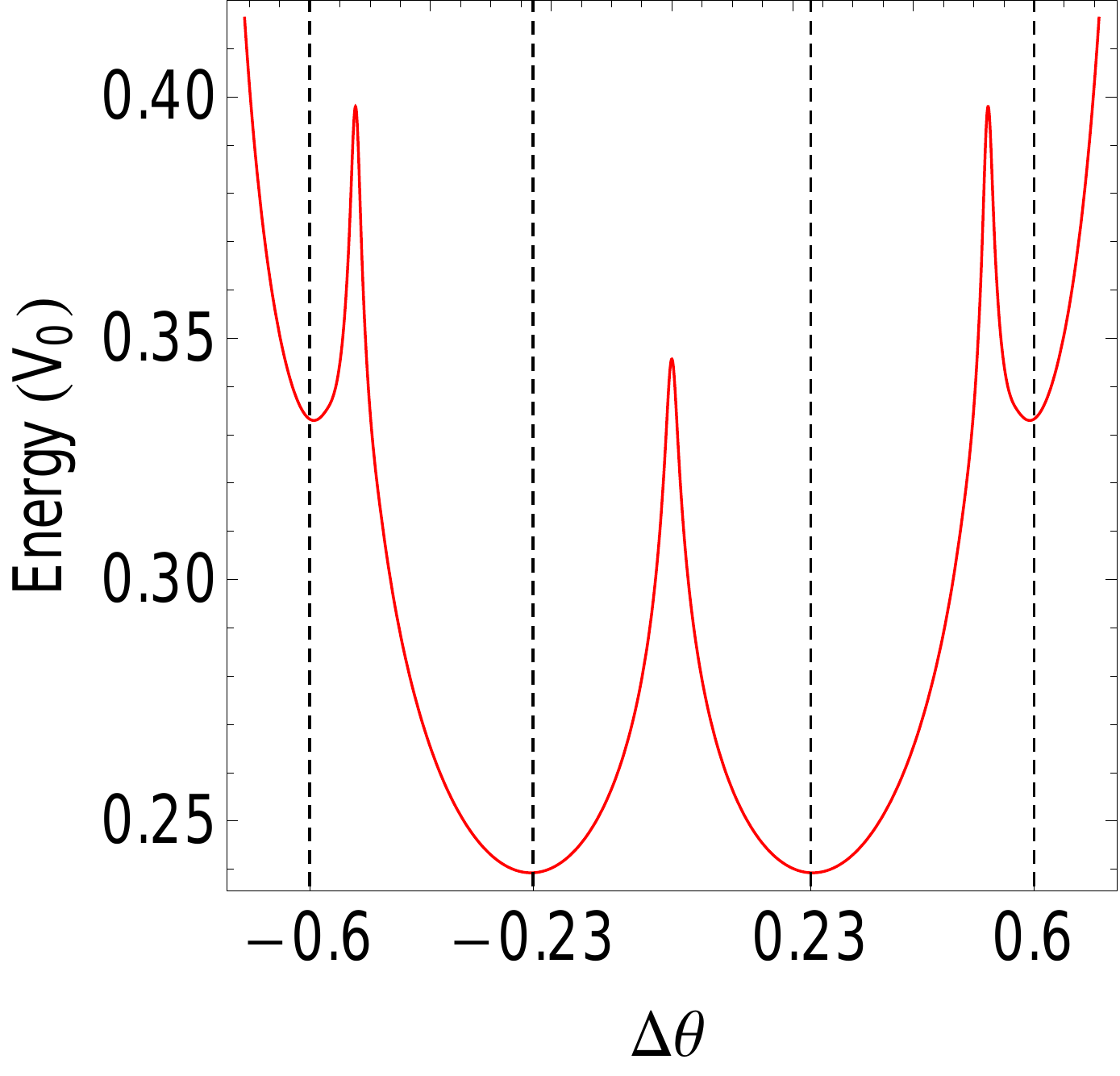}

\caption{Helical solitons  separating helices of different winding angle. The string is shown in three dimension (top), in cylindrical coordinates (second panel) and in repeated cylindrical coordinates (third panel) which illustrates the curve $ {\boldsymbol \psi}(s)$. At the bottom  the  energy of an helical structures as a function of its gain angle $\Delta \theta$ (radiants) between consecutive monomers, for two different choices of $\sigma$,  with (right) and without (right) metastable states.}
\label{draw}
\end{figure}

Before proceeding we  motivate M1 by introducing M2, a more faithful ball-and-spring model  of  non-locally interacting monomers of cylindrical coordinates $\theta_i$, $z_i$, which we use in numerical simulations.  Monomers $i$ and $i+1$ interact harmonically via $K (d_{i,i+1}-a)^2/2$ ($d_{ij}$ is their distance) so that the chain is  floppy, as for ssDNA~\footnote{Clearly, $a$ is in general slightly smaller than the actual equilibrium length of a straight polymer, because of the electrostatic repulsion}. They also interact repulsively via a screened coulomb potential $U_{ij}= \sigma_{ij} U_o \exp({-d_{ij}/d_o})/d_{ij}$. The modulation factor $\sigma_{ij}=\sigma(\theta_{ij})$ reflects the cylindrical nature of the screening from the tube as well as possible effects of adhesive optimization well known in the case of ssDNA-nanutobe hybrids~\cite{Manohar}. In absence of a corrugation potential we choose a sufficiently smooth function of period $\pi$, $\sigma_1(\theta)=[1 + \cos(\theta/2)^2]/2$. To allow for the existence of metastable states we also consider $\sigma_2=\left[1+ \cos(6 \theta)^2 \right]/2$. More parametrized choices might be needed to faithfully address  specific situations, yet they do not  qualitatively change our results. In simulations we  choose $r=9$, $a=7$, $d_o=100$, $V_o=10$, $K=1$, which corresponds, if lengths are measured in \AA, to charged ssDNA on a nanotube of  diameter of 1.8 nm, with a Debye screening length of $10$ nm. We choose  $V_o/K=10$ to ensures a electrostatic stretch of less than 10\% of $a$. The actual value of $V_o$ simply defines the timescale (in ratio $V_o/m$ with the mass $m$ of the monomer).

Figure 1, bottom left, shows the double-dip shape of the total energy of M2 when restricted to an helical configuration, as a function of the wrapping angle, when we choose $\sigma=\sigma_1$ as screening function. Physically, the two opposite stable angles (which depend on $r/a$) come from a competition: winding the helix increases the screening, but also the repulsion between monomers whose distance is shortened. Now we can justify the locality of M1 (M2 is obviously non-local). Because we study low energy dynamics on helical manifolds separated by kinks we approximate the energy with the last two  terms in Eq. (\ref{L}):%, local in derivatives of $ {\boldsymbol \psi}(s,t)$
 one is the energy of the helix (Fig. 1), which depends on its angle (and thus $ \boldsymbol T$); then self-repulsion  provides and extra effective bending rigidity. The agreement between analytical solutions of M1 and numerics on M2 confirms the choice. Figure 1, bottom right shows the helical energy in the case of $\sigma_2$, demonstrating the existence of metastable configurations.  

(Quasi)degeneracy implies kinks between (meta)stable structures. Before investigating numerically their Newtonian dynamics we gain theoretical insight by solving M1. The equations of motion from $M1$ for a string of length $2l$  are obtained by minimizing the  constrained Lagrangian 
\begin{equation}
L=\int_{-l}^{l}\left[{\cal L}-\frac{1}{2}\mu\left( {\boldsymbol T}^2-1\right)\right] \mathrm{d}s +{\boldsymbol{F}_{\!\!+} \!\! \cdot {\boldsymbol \psi}}(l)-{\boldsymbol{F}_{\!\!-}}  \cdot {\boldsymbol \psi}(-l),
\label{LL}
\end{equation}
where $\mu(s)$ is a functional Lagrange multiplier ensuring ${\boldsymbol T}^2=1$, while ${\boldsymbol F}_{\! \! +}$ is the force  exerted at the boundary  ${\boldsymbol\psi}(+ l )$ and $-{\boldsymbol F}_{\! \! -}$  at ${\boldsymbol\psi}(-l )$. This returns the  problem
\begin{eqnarray}
\lambda \ddot{{\boldsymbol \psi}}&=&-\partial_s {\boldsymbol j} \nonumber \\
{\boldsymbol j}_{|_{\pm l}}&=&-{\boldsymbol F_{\!\! \pm}},
\label{motion}
\end{eqnarray}  
 a  conservation equation for the density of momentum $\lambda \dot{{\boldsymbol \psi}}$, of  flux    
\begin{equation}
{\boldsymbol j}=-\partial_{\boldsymbol T} V+k \partial^2_{s} {\boldsymbol T} -\mu {\boldsymbol T}.
\label{flux}
\end{equation}

If $V$ has local minimum in $\bar {\boldsymbol T}$, then an helix  ${\boldsymbol \psi}(s) = \bar {\boldsymbol T}s$ is a static solution. It exists when the  forces applied at the extremities are purely tensile and balanced, ${\boldsymbol F_{\pm}}=\bar{ \boldsymbol T} F$. Since ${\boldsymbol j}$ is  the stress vector of our 1D system,  Eq. (\ref{flux}) shows that  $\mu(s)$ is  the scalar tensile stress (which for a stable helix is from Eq. (\ref{flux})  the only stress  $\mu(s)=\mathrm{const}=F$)~\footnote{Clearly ${\boldsymbol \psi}(s) = \bar {\boldsymbol T}s+{\boldsymbol w}t$ would also be a solution, corresponding to a translating/rotating helix.}. 

\begin{figure}[t!]   
\includegraphics[width=2.8 in]{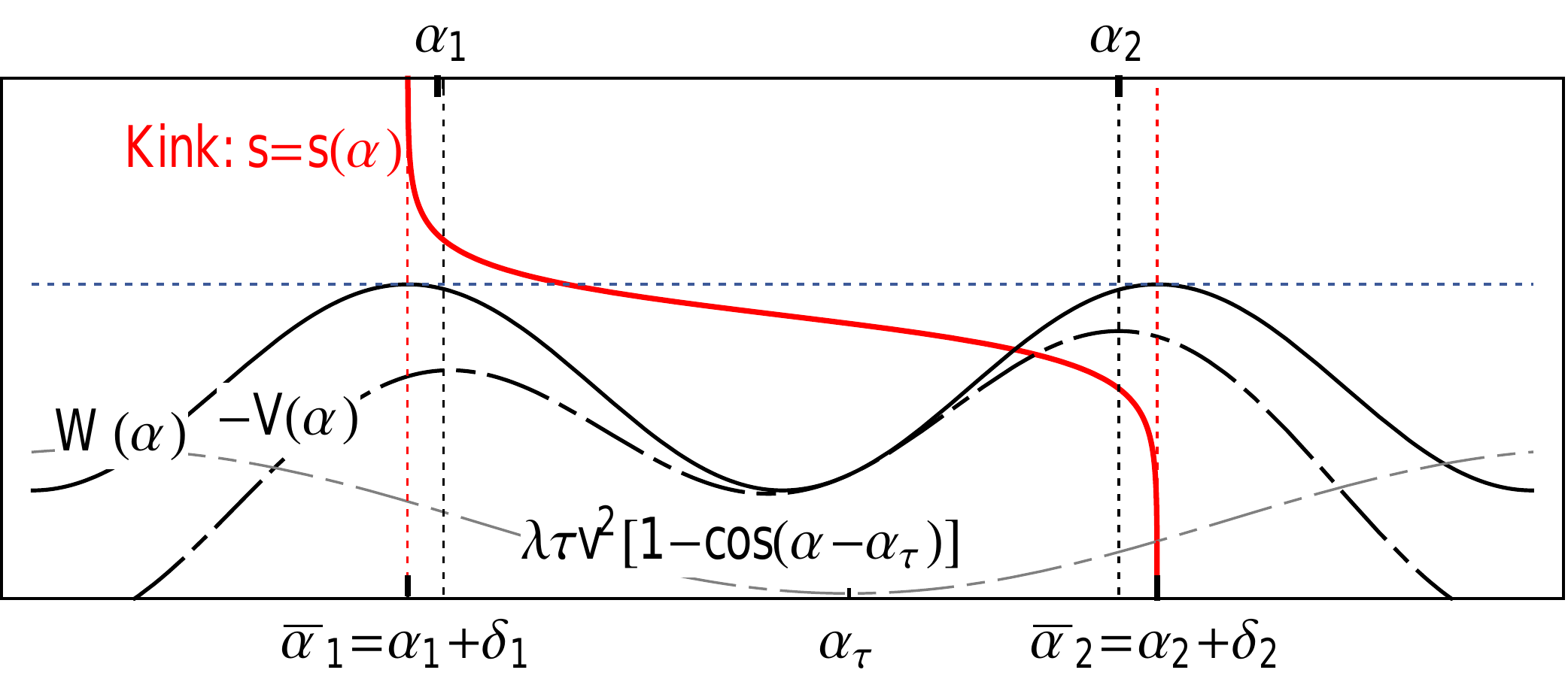}
\includegraphics[width=2.8 in]{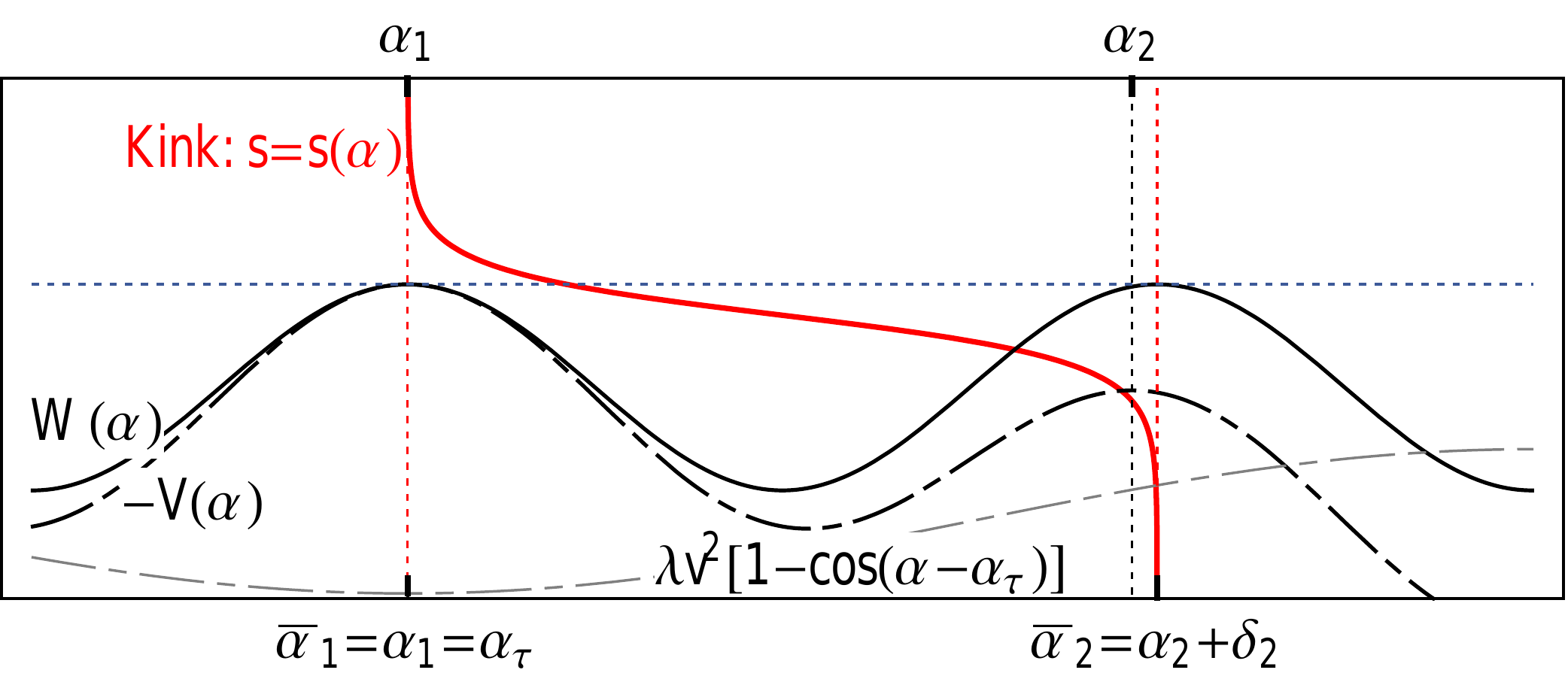}

\caption{Schematics of a solution of Eq. (\ref{motionalpha}). Top: in general a soliton correspond to a trajectory connecting two maxima of $W$ (solid black line). From Eq. (\ref{W}), maxima of $W$ do not correspond to minima  
of $V$ (dashed black line), due to the extra term $\lambda \tau v^2 \left[1-\cos(\alpha-\alpha_{\tau})\right]$ (dashed grey line). It follows that solitons between degenerate structures are possible but they must have a specific speed. Bottom: the solution for a soliton between helices of different energy $V$, the same obtained by simulations shown in Fig.~3 (see  Sup. Mat. S1).
}
\label{draw}
\end{figure}
%
%In the same Fig.~1 we plot a metastable configuration, obtained through over-damped dynamics of M2 starting from a straight configuration parallel to the nanotube axis. It consists, as expected,  of alternating helices of opposite angle, separated by dense U-kinks. 

Equations (\ref{motion}) show that an helical structure can  change its pitch via uniform compression/expansion. Indeed an uniform rotation of the tangent vector ${\boldsymbol T}=e^{i \omega t}$ (in the complex plane representation of  2D vectors) and therefore ${\boldsymbol \psi}(s,t)= {\boldsymbol T}s +{\boldsymbol w}t$, are a solution. Substituting into Eq. (\ref{motion}) we obtain the tension $\mu(s)=\mu_0-s^2 \omega^2/2\lambda$, where $\mu_0$ is a constant which depends on the forces applied at the boundaries: from Eq. (\ref{motion}) we have for the tangentially applied forces at the boundaries ${\boldsymbol T}_{\pm} \cdot {\boldsymbol F}_{\pm}=\mu_0 -l^2 \omega^2/\lambda$, whereas  the normally applied forces account for the needed torque:  ${\boldsymbol N}_{\pm} \cdot {\boldsymbol F}_{\pm}= \partial_{{\boldsymbol T}} V$ (where ${\boldsymbol N}={\boldsymbol T}'$). This solution  is problematic as   stresses diverge with  size and so does  speed ( $\dot{\boldsymbol \psi}_{\pm}=\pm l\omega  {\boldsymbol N}+{\boldsymbol w}$): it is thus only viable for a finite structure, with properly applied loads.

\begin{figure}[t!]   
\begin{center}
\includegraphics[width=2.5 in]{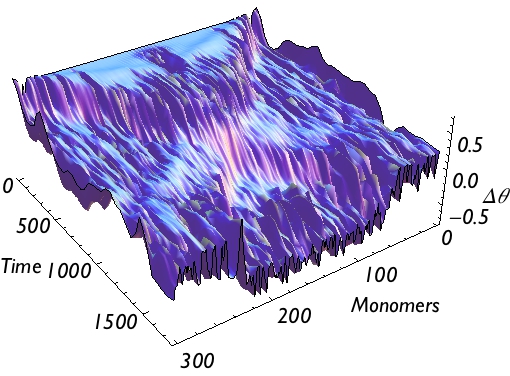}\vspace{5mm}
\includegraphics[width=1.55 in]{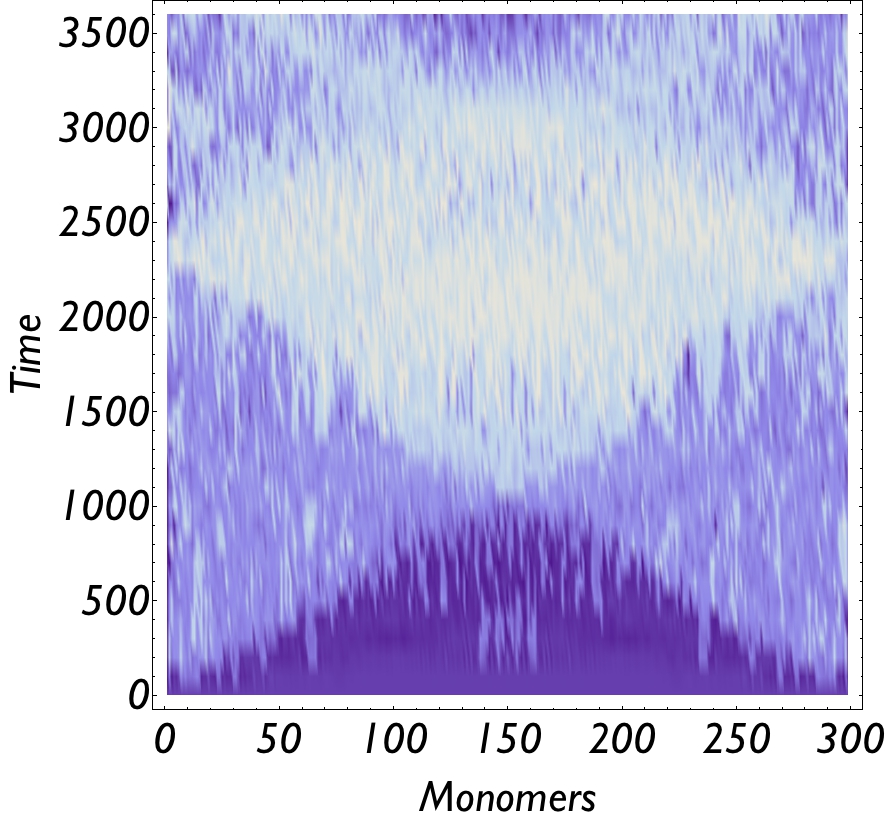}\hspace{3mm}\includegraphics[width=1.5 in]{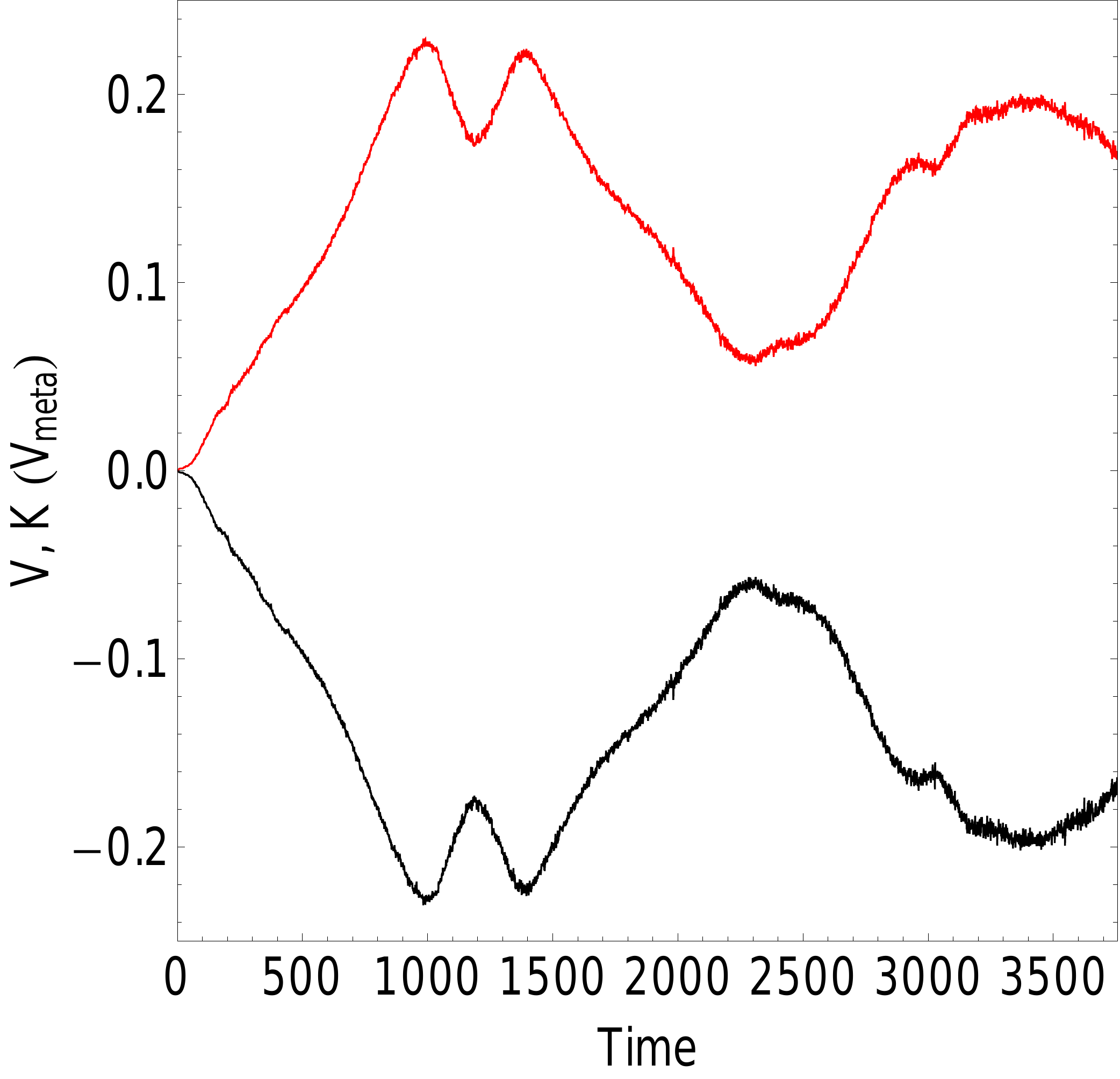}
 \end{center}
\caption{Helical solitons  separating helices of different winding angle, propagating  into a region of higher potential energy (numerical integration of the Newtonian dynamics of M2,  animation in  supp. mat. S1). The system starts in a metastable helical configuration (Fig1 bottom left panel). Stable helical configurations forms at the boundaries and propagates inside. Top: 3D plot for the angular deviation $\Delta \theta_i=\theta_{i+1}-\theta_i$ vs. space and time. Bottom left: density plot for the angular deviation $\Delta \theta_i=\theta_{i+1}-\theta_i$ (also plotted in 3D at the bottom) as a function of time and monomers, demonstrating propagation at fixed speed, collision, and reflection on the boundaries. Bottom, right: the time evolution of kinetic (red) and potential (black) energy ($V_m$ is the  energy of the initial metastable configuration) demonstrates the  expected  initial linear growth of kinetic energy until collision; after collision a stable helix of opposite orientation forms with solitons now propagating outward until reflection.}
\label{draw}
\end{figure}

However an helical structure can change its pitch without divergences in velocities   by propagating a soliton. A traveling solution of Eqs (\ref{motion}), (\ref{flux}) has  the form ${\boldsymbol \psi}={\boldsymbol \phi}(s-vt)+{\boldsymbol w} t$, which implies ${\boldsymbol T}={\boldsymbol \phi}'$. Then Eq. (\ref{motion}) becomes $\lambda v^2 {\boldsymbol T}'=-{\boldsymbol j}'$ which can be integrated to obtain
\begin{equation}
k{\boldsymbol T}''=-\partial_{\boldsymbol T} W({\boldsymbol T})+\mu {\boldsymbol T}.
\label{motionsphere}
\end{equation}
Eq. (\ref{motionsphere}) is simply a Newton equation for a ``particle'' described by ${\boldsymbol T}(s)$ (where now $s$ is ``time'') constrained to a circumference and subject to the potential
\begin{eqnarray}
 W({\boldsymbol T})&=&-V({\boldsymbol T})+\frac{1}{2}\lambda v^2({\boldsymbol T} -{\boldsymbol \tau})^2, 
\label{W}
\end{eqnarray}
where ${\boldsymbol \tau}$ is defined by 
\begin{eqnarray}
 { \boldsymbol F}_{\! \!-} =\lambda v^2
({\boldsymbol T}_{-}-{\boldsymbol \tau})
\end{eqnarray}
and ${\boldsymbol T}_{\pm}={\boldsymbol T}(\pm l)$. Equation~(\ref{motionsphere}) becomes more manageable if projected on its Frennet-Serret frame~\cite{Serret}. We define $\alpha$  via {${\boldsymbol T}= e^{i\alpha}$}, and similarly  {${\boldsymbol \tau}= \tau e^{i\alpha_{\tau}}$}. Then projection on the normal vector ${\boldsymbol N}={\boldsymbol T}'$  yields finally
\begin{equation}
k\alpha''=-\partial_{\alpha}W(\alpha),
\label{motionalpha}
\end{equation}
a simple 1D Newton equation for a particle in  potential
\begin{equation}
W(\alpha)=-V(\alpha)+\lambda \tau v^2 \left[1-\cos(\alpha-\alpha_{\tau})\right].
\label{W2}
\end{equation}
Also, projection on the tangent ${\boldsymbol T}$ gives the tensile stress
\begin{equation}
\mu=\lambda v^2(1-{\boldsymbol \tau}\cdot {\boldsymbol T})-k\alpha'^2.
\label{motionmu}
\end{equation}

From Eqs (\ref{motionalpha}), (\ref{motionmu}) the phonon dispersion in a stable helix $\bar \alpha$ is found to be $\omega^2=c^2_Fq^2+(k/\lambda)q^4$, with $c_F^2=c^2+F/\lambda$ the tension-modified speed of sound, and $c$ the speed of sound in absence of tension, $c^2=\partial^2_{\alpha}V|_{\bar \alpha}/\lambda$. Note that in absence of a direction-dependent potential $c=0$ and we regain the usual dispersion  for a string under tension. Not also that the helical structure is  stable to applied pressure ($F<0$) when it does not exceed the critical value $-F>\lambda c^2$.

We are now in familiar territory. If we consider $l\to \infty$ then a topological soliton connecting two different helical structures corresponds to a trajectory between two {\it degenerate} maxima of $W$~\cite{Lubensky, Dauxois}. These might come from {\it non-degerate} local minima of $V$. 

When $v=0$, and the kink is static, from Eq.~(\ref{W}) $W=-V$ and  maxima of $W$ are minima of $V$ and correspond to stable helices: static kinks are thus only possible between {\it degenerate} structures of the same energy, and never between stable and metastable structures~\footnote{Unless of course proper forces are applied at the end, thus changing the energetics.}. 

Most interestingly,  however,  Eq.~(\ref{W}) and Fig.~2 (top panel) show that  the extra term in $W$, proportional to the square of the speed implies that even if  two helices do not have the same energy,  a soliton can still exist between them (much unlike the sine-Gordon case),  {\it but it must move, and with a locked speed}. Indeed only $v\ne0$ in the second term of Eq. (\ref{W}) can make the effective potential $W$ degenerate when $V$ is not. 

The physical reason for this mechanism is rather intuitive. The propagation of a soliton corresponds to an  homotopy between continuum states of different topological invariant (winding angle) per unit length. This constrains a rotation of one domain with respect to the other. Consider a soliton propagating inside a static region (2) of higher energy, leaving a helix of lower energy (1) in its wake. Because of continuity, helix B must rotate with respect to A, with speed $\dot{\boldsymbol  \psi}_1=v({\boldsymbol T}_2-{\boldsymbol T}_1)$. As the soliton propagates at constant speed $v$, the total  kinetic energy increases linearly in time with rate $ \lambda v^3({\boldsymbol T}_1-{\boldsymbol T}_2)^2/2$ while the potential energy decreases linearly in time with rate $(V_2-V_1) v$. Then energy conservation  locks the speed of the soliton at
\begin{equation}
v^2=2\frac{V_1-V_2}{\lambda ({\boldsymbol T}_1-{\boldsymbol T}_2)^2}.
\label{v}
\end{equation}

Remarkably, this heuristic  formula precisely returns the speed $v$ that makes $W$ of Eq. (\ref{W}) degenerate, (having chosen ${\boldsymbol \tau}={\boldsymbol T}_1$, or ${\boldsymbol F}_{\!\!-}=0$). This can be seen clearly in Fig.~2, bottom panel, which predicts the existence of a soliton of speed $v$ given by (\ref{v}) between non degenerate (meta)stable helices. It clarifies that in these system energy and momentum are not localized inside the soliton (as in a sine-Gordon case), but rather flow through the soliton as it propagates. This can also be understood from  Eq. (\ref{motion}) from which we have  ${\boldsymbol j}=\lambda v^2 ({\boldsymbol T_1- \boldsymbol T})$: the flux of momentum  is uniform in the helical structures but changes through the soliton.

\begin{figure}[t!]   
\begin{center}
\includegraphics[width=2.8 in]{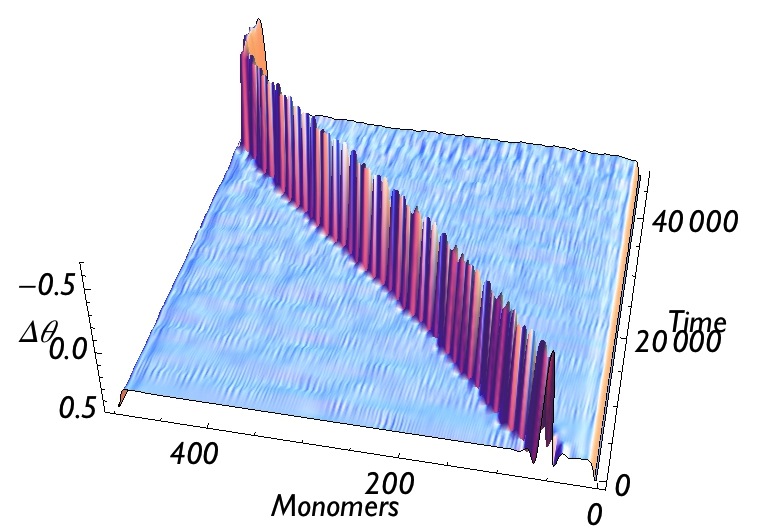}
\includegraphics[width=3.3 in]{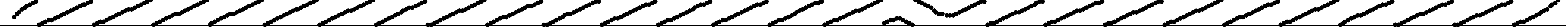}
\includegraphics[width=3.3 in]{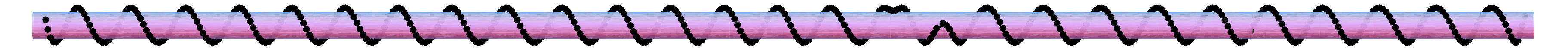}
\includegraphics[width=3.3 in]{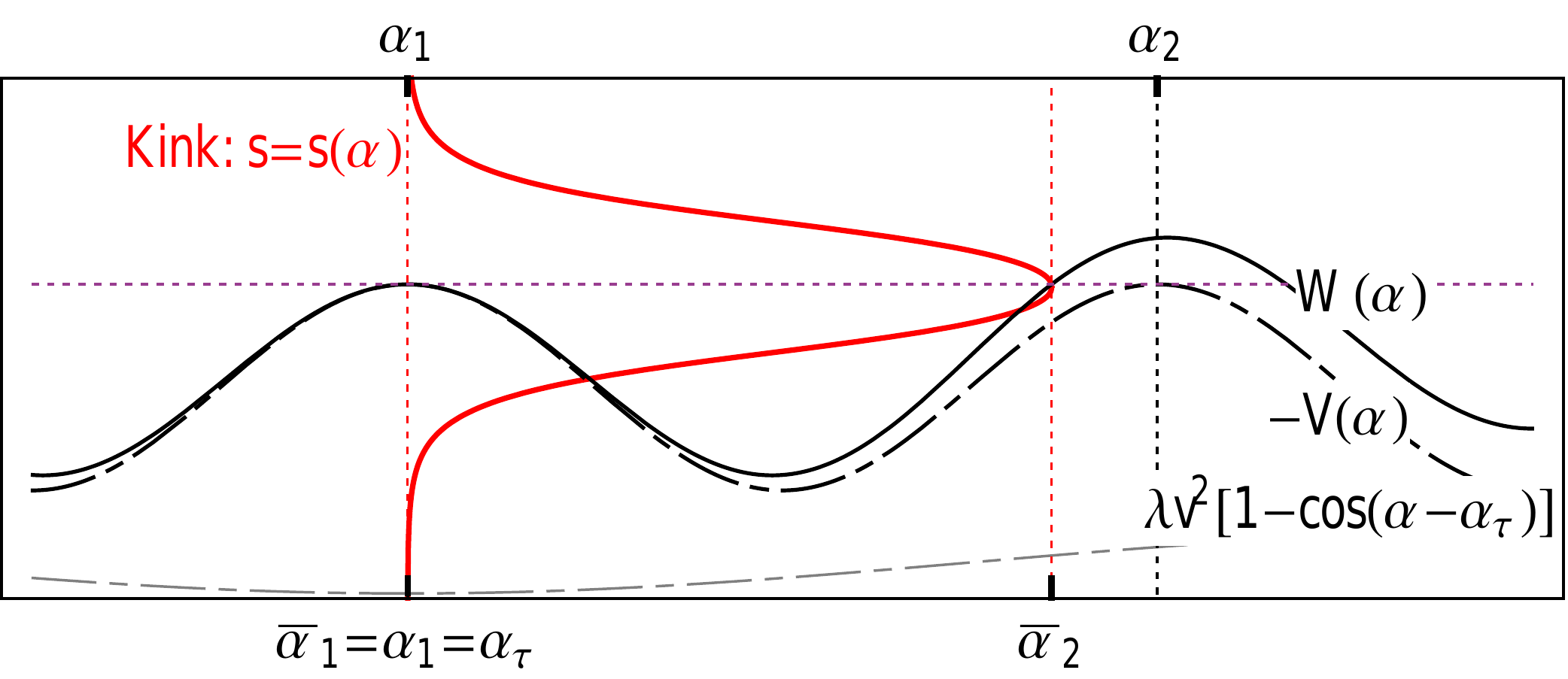}

 \end{center}
\caption{Pulse soliton propagating at uniform speed, obtained by velocity-Verlet integration of M2. Top: Plot of the angular deviation $\Delta \theta_i=\theta_{i+1}-\theta_i$ between consecutive monomers as function of time. Below we show a snapshot of the pulse in cylindrical coordinates and in 3D. A pulse soliton is predicted by our analytical framework M1 (bottom panel) as the speed $v$ raises the maximum in $V(\alpha_2)$, allowing for a trajectory that bounces back from $\alpha_2$ and returns to $\alpha_1$. As $ v\to 0$ the size of the pulse increases ultimately tending to two static kinks places infinitely far away. 
}
\label{draw}
\end{figure}

We use M2 to corroborate this result. Figure 3 shows results of velocity-Verlet numerical integration of M2. An helix is prepared in a metastable state corresponding to $\Delta \theta \simeq 0.6$~rad (Fig. 1 bottom right panel), with open boundaries. Lower energy helices ($\Delta \theta \simeq 0.23$~rad) form at the boundaries and propagate inside with constant speed, as the potential energy decreases linearly, and the kinetic energy correspondingly increases. We see from the simulation that upon collision a new metstble domain ($\Delta \theta \simeq -0.6$ rad) forms  and the potential energy starts increasing again, until reflection with the boundaries. As the simulation proceeds more energy of the solitonic dynamics is dissipated into phonons, as expected in a discrete system (see supp. mat. S1). 

We see that the low-energy physics  still affords a description in terms of excitation dynamics (kinks), by folding the domain dynamics into a velocity dependent effective potential, which however fixes the speed of the kinks. However the union of a kink-antikink separates two identical domains and can thus--at least in principle--propagate at any speed. 

Such pulses are admitted by our analytical framework M1. Consider for instance two degenerate minima of V, $\alpha_1$ and $\alpha_2$, as in Fig.~4. We choose ${\boldsymbol F}_{\!\!-}=0$ and thus
 ${\boldsymbol \tau}={\boldsymbol T}_-={\boldsymbol T}_1$. Now $W$ has still a maximum in $\alpha_1$, but an higher maximum in $\alpha_2+\delta$ slightly shifted from $\alpha_2$. A trajectory can now start in $\alpha_1$, reach the proximity of the new structure, and then come back to $\alpha_1$, thus describing a kink-antikink pair (Fig.~4, bottom panel). Clearly there is an upper limit for $v$ given by the speed of sound of the helical structure, previously defined: when $v>c$, $\alpha_1$ becomes a minimum of $W$ and no solitonic solution is possible.   In Fig.~4 we show results of simulations on M2 demonstrating stability and motion of such  pulses (see supp. mat. S2).  
 
Note  that as $v$ goes to zero, and $W(\alpha_2)$ becomes degenerate with $W(\alpha_1)$,  the trajectory in Fig.~4 (bottom) would describe two opposite kinks progressively far away from each other.  It is not difficult (details will be shown elsewhere~\cite{NisoliFuture}) to compute the total energy for a traveling pulse of speed $v$ and obtain that in the limit of low speed the energy decreases tending for $v \to 0$ to the energy of two static kinks, placed infinitely far away and thus non-interacting. Since in a discrete system soliton propagation is associated with phonon radiation one can imagine that a pulse will always decay into two distant static kinks. However we see easily that topology again protects from this dynamics: increasing the size of the pulse must force rotation on the domain $\alpha_1$, which is assumed infinitely long, at infinite cost of kinetic energy. For finite systems however the decay is possible: indeed simulations of exactly the same situation depicted in Fig. 4, but on 300 rather than 500 monomers, show such decay (supp. mat. S3), as the energy needed to set the external domain into rotation is  inferior to the energy stored in the bound kink-antikink. 
 
In conclusion we have  presented the evolution of  topological solitons between helical structures of different winding angle for unit length. These solitons are stable but different from  e.g. sine-Gordon-like solitons, as their velocities are are controlled by the competition between kinetic energy of the domain motion and and steady  changes in potential energy due to helical wrapping/unwrapping.

We are grateful to S. Kilina and D. Yarotski for useful discussions. The work  was supported by  US DOE BES E304, KAW  and ERC DM 321031,  under the auspices of the National Nuclear Security Administration of the US DOE at Los Alamos National Laboratory under Contract No. DEAC52-06NA25396.

\end{document}